\documentclass[sigconf]{acmart}
\usepackage{balance}
\newcommand{\revise}[1]{\textcolor{black}{#1}}
\usepackage{amsmath,amsfonts}

\usepackage{graphicx}
\usepackage{textcomp}

\usepackage{booktabs}
\usepackage{color, colortbl}
\usepackage{caption}
\usepackage{subcaption}
\usepackage{amsbsy}
\usepackage{pifont} 
\usepackage{multirow}
\usepackage[normalem]{ulem}
\usepackage{array}
\hypersetup{
    colorlinks=true,
    linkcolor=blue,
    citecolor=red
}
\usepackage{bbding}
\usepackage[super]{nth}
\usepackage{soul}

\definecolor{2nd}{gray}{0.8}

\newcolumntype{C}{!{\vrule width 1.1pt}c}
\newcolumntype{g}{>{\columncolor{Gray}}c}

\newcolumntype{^}{>{\currentrowstyle}}

\newcommand{\redtick}{\textcolor{red}{\ding{52}}}

\newcolumntype{Y}{>{\centering\arraybackslash}X}
\newcolumntype{P}[1]{>{\centering\arraybackslash}p{#1}}

\definecolor{LightCyan}{rgb}{0.88,1,1}

\definecolor{mypurple}{RGB}{153, 0, 153}
\definecolor{mygray}{RGB}{128, 128, 128}
\definecolor{mygreen}{RGB}{0, 153, 0}

\definecolor{mycyan}{RGB}{64, 128, 128}
\definecolor{mypink}{RGB}{255, 182, 193}
\definecolor{myred}{RGB}{165,42,42}
\definecolor{myyellow}{RGB}{255, 191, 0}

\definecolor{tab_red}{rgb}{0.71, 0.11, 0.0}
\definecolor{tab_green}{rgb}{0.11, 0.71, 0.0}

\newcommand{\thickhline}{%
	\noalign {\ifnum 0=`}\fi \hrule height 1pt
	\futurelet \reserved@a \@xhline
}
\makeatletter
\global\let\oriCT@@do@color\CT@@do@color

\usepackage{gensymb}

\usepackage{textcomp}

\usepackage{enumitem}
\definecolor{custompurple}{HTML}{6D2382}
\definecolor{urlblue}{HTML}{005698}

\usepackage{algorithm}
\usepackage{algpseudocode}
\usepackage{amsmath}
\usepackage{algpseudocode}
\newcommand{\func}[1]{\mathsf{#1}}

\algnewcommand{\nComment}[1]{\Statex \Comment{#1}}
\algdef{SE}[SUBALG]{Indent}{EndIndent}{}{\algorithmicend\ }%
\algtext*{Indent}
\algtext*{EndIndent}

\AtBeginDocument{%
  }

\copyrightyear{2025}
\acmYear{2025}
\setcopyright{acmlicensed}
\acmConference[MM '25] {Proceedings of the 33rd ACM International Conference on Multimedia}{October 27--31, 2025}{Dublin, Ireland.}
\acmBooktitle{Proceedings of the 33rd ACM International Conference on Multimedia (MM '25), October 27--31, 2025, Dublin, Ireland}
\acmISBN{979-8-4007-2035-2/2025/10}
\acmDOI{10.1145/3746027.3754919}

\definecolor{ForestGreen}{RGB}{34,139,34}

\begin{document}

\title{CCStereo: Audio-Visual Contextual and Contrastive Learning for Binaural Audio Generation}

\author{Yuanhong Chen}
\authornote{\revise{Work done during a research internship at Sony AI.}}
\email{yuanhong.chen@adelaide.edu.au}
\affiliation{%
  \institution{Australian Institute for Machine Learning, University of Adelaide}
  \city{Adelaide}
  \country{Australia}
}

\author{Kazuki Shimada}
\email{kazuki.shimada@sony.com}
\affiliation{%
  \institution{Sony AI}
  \city{Tokyo}
  \country{Japan}
}

\author{Christian Simon}
\email{christian.simon@sony.com}
\affiliation{%
  \institution{Sony Group Corporation}
  \city{Tokyo}
  \country{Japan}
}

\author{Yukara Ikemiya}
\email{yukara.ikemiya@sony.com}
\affiliation{%
  \institution{Sony AI}
  \city{Tokyo}
  \country{Japan}
}

\author{Takashi Shibuya}
\email{takashi.tak.shibuya@sony.com}
\affiliation{%
  \institution{Sony AI}
  \city{Tokyo}
  \country{Japan}
}

\author{Yuki Mitsufuji}
\email{yuki.mitsufuji@sony.com}
\affiliation{%
  \institution{Sony AI, Sony Group Corporation}
  \city{New York}
  \country{USA}
}



\begin{abstract}
Binaural audio generation (BAG) aims to convert monaural audio to stereo audio using visual prompts, requiring a deep understanding of spatial and semantic information. The success of the BAG systems depends on the effectiveness of cross-modal reasoning and spatial understanding. Current methods have explored the use of visual information as guidance for binaural audio generation. However, they rely solely on cross-attention mechanisms to guide the generation process and under-utilise the temporal and spatial information in video data during training and inference. These limitations result in the loss of fine-grained spatial details and risk overfitting to specific environments, ultimately constraining model performance. In this paper, we address the aforementioned issues by introducing a new audio-visual binaural generation model with an audio-visual conditional normalisation layer that dynamically aligns the target difference audio features using visual context. To enhance spatial sensitivity, we also introduce a contrastive learning method that mines negatives from shuffled visual features. We also introduce a cost-efficient way to utilise test-time augmentation in video data to enhance performance. Our approach achieves state-of-the-art generation accuracy on the FAIR-Play, MUSIC-Stereo, and YT-MUSIC benchmarks. Code is available at \textcolor{urlblue}{\textit{\url{https://github.com/SonyResearch/CCStereo}}}.
\end{abstract}




\maketitle

\section{Introduction}
\label{sec:intro}


Binaural audio is gaining significant attention in streaming media, revolutionising how listeners experience sound in a digital environment. This technology finds applications in various domains, including virtual reality (VR) \cite{hoeg2017binaural}, 360-degree videos \cite{morgado2018self}, and music \cite{griesinger1990binaural}.
By simulating a two-dimensional soundscape, binaural audio creates a deeply immersive experience, allowing listeners to feel as if they are physically present within the auditory scene.


\begin{figure}
    \centering
    \includegraphics[width=.95\linewidth]{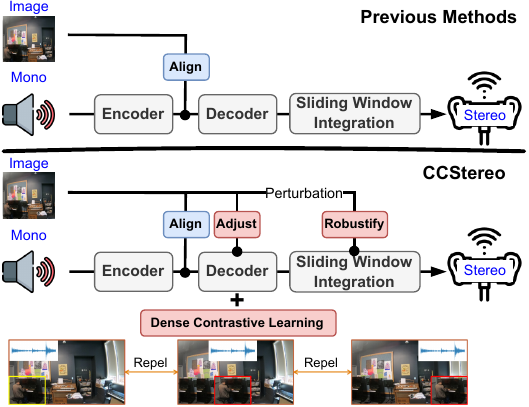}
    \vspace{-12pt}
    \caption{
    Comparison between previous mono-to-binaural methods~\cite{gao20192,zhou2020sep} (top) and our proposed CCStereo framework (bottom). While prior approaches rely on implicit global alignment, CCStereo explicitly targets three key aspects of the spatialisation process: (1) \textbf{align} establishes audio-visual correspondence; (2) \textbf{adjust} the predicted stereo features by matching the mean and variance of the target; and (3) applying visual perturbations during training and inference to robustify to the prediction. In addition, CCStereo incorporates dense contrastive learning to improve spatial sensitivity through discriminative supervision across visual contexts.
    }
    \vspace{-15pt}
    \label{fig:motivation}
    \Description[Comparison between previous methods and our method.]{Comparison between previous methods and our method.}
\end{figure}

Binaural audio recording typically requires specialised hardware like dummy head systems~\cite{gao20192}. These systems are costly and lack portability, making them impractical for everyday use. To address this, researchers have developed methods to spatialise audio from monaural recordings, known as binaural audio generation (BAG)~\cite{gao20192,zhou2020sep,xu2021visually}. These methods use visual information to estimate the differential audio between left and right channels. However, existing frameworks often rely on simple feature fusion strategies, which may struggle to capture complex visual-spatial relationships, limiting their generalisability and performance.
To better utilise the visual information, previous works~\cite{gao20192,zhou2020sep,xu2021visually,parida2022beyond,li2024cross,li2024cyclic} have explored various strategies to enhance semantic and spatial awareness across modalities. These approaches aim to improve cross-modal feature interaction~\cite{zhou2020sep,xu2021visually,parida2022beyond,zhang2021multi}, strengthen spatial understanding~\cite{garg2021geometry,li2024cross,li2024cyclic}, and incorporate 3D environmental cues~\cite{garg2021geometry}.
However, these methods still rely on concatenation or cross-attention to guide the generation process. While cross-attention excels in blending features from different modalities (i.e., representation fusion~\cite{xu2013survey,li2018survey,baltruvsaitis2018multimodal,xu2021visually, xuan2020cross}), it is weak at aligning and maintaining spatial fidelity in the audio, making it less effective for integrating fine-grained conditioning information.

In addition, existing models remain prone to overfitting the training environment due to their reliance on specific data distributions and insufficient regularisation mechanisms. These issues often result in limited generalisation to diverse or unseen scenarios.
Unfortunately, the structure of the widely used FAIR-Play~\cite{gao20192} dataset fails to address this concern, as a significant amount of scene overlap has been observed between the training and testing sets~\cite{xu2021visually}, resulting in overly optimistic evaluation results on the current benchmark. Xu et al.~\cite{xu2021visually} tackled this issue by reorganising the dataset based on clustering results of scene similarity. 
Additionally, methods involving training on synthetic stereophonic data from external sources~\cite{zhou2020sep, xu2021visually} and incorporating depth estimation~\cite{parida2022beyond} have also shown potential in mitigating the overfitting problem.
Despite promising results, these approaches rely on additional single-source audio data for synthetic training, introducing extra cost and complexity. 
They also under-utilise the inherent temporal and spatial information in video data at both training and inference time, missing the opportunities to improve prediction robustness and spatial consistency.

In this paper, we introduce a novel U-Net-based generation framework, named as \underline{C}ontextual and \underline{C}ontrastive \underline{Stereo}phonic Learning (CCStereo), which aims to address the aforementioned challenges.
The framework consists of a visually adaptive stereophonic learning method that enhances cross-modal ``alignment'' and enables ``adjustment'' to the generation process based on the provided spatial information, along with a robustified and cost-effective inference strategy, as shown in Fig.~\ref{fig:motivation}.
Unlike previous methods that rely solely on concatenated~\cite{gao20192,zhou2020sep,xu2021visually,li2024cross,liu2024visually} or cross-attended~\cite{zhang2021multi} features for differential audio generation, we adopted the concept of conditional normalisation layers~\cite{huang2017arbitrary,nam2018batch} from image synthesis field to control the generation process through estimated mean and variance shifts informed by visual context. 
Additionally, we propose a novel audio-visual contrastive learning method that improves the model's spatial sensitivity by enforcing feature discrimination across the anchor frame, nearby frames, and the spatially shuffled anchor frame. This encourages the model to learn more discriminative representations of different object locations and their corresponding generated spatial audio, as illustrated by the simulated position change of the piano in Fig.~\ref{fig:motivation}.
Moreover, the widely used sliding window inference strategy~\cite{gao20192} introduces significant redundancy due to substantial frame overlap, which is common in video data. We argue that this overlap presents an opportunity to adopt test-time augmentation (TTA), leveraging the redundant information to enhance robustness and improve prediction accuracy. We introduce Test-time Dynamic Scene Simulation (TDSS), which divides the video into $N$ sets of five consecutive frames and applies five-crop augmentation to each set across the entire video.
To summarise, our main contributions are
\begin{itemize}[topsep=0pt]
    \item An audio-visual conditional normalisation layer that adjusts feature statistics based on visual context to enhance spatial control in difference audio decoding process.
    \item A novel audio-visual contrastive learning method that enhances spatial sensitivity by mining negative samples from nearby frames and spatially shuffled visual features to simulate object position changes.
    \item A cost-efficient \textit{Test-time Dynamic Scene Simulation (TDSS)} strategy that exploits frame redundancy from sliding window inference by applying five-crop augmentation to consecutive frame sets for improved robustness and accuracy.
\end{itemize}
We demonstrate the effectiveness of our CCStereo model
on established benchmarks, including the FAIR-Play dataset~\cite{gao20192} with both the original 10-split~\cite{gao20192} and the more challenging 5-split protocols~\cite{xu2021visually}. 
Additionally, we extend our evaluation to two real-world datasets, MUSIC-Stereo~\cite{xu2021visually} and YT-MUSIC~\cite{morgado2018self}, demonstrating better generalisation across diverse audio-visual scenarios and superior generation quality with an efficient architecture.

\begin{figure*}[ht]
    \centering
    \includegraphics[width=0.92\textwidth]{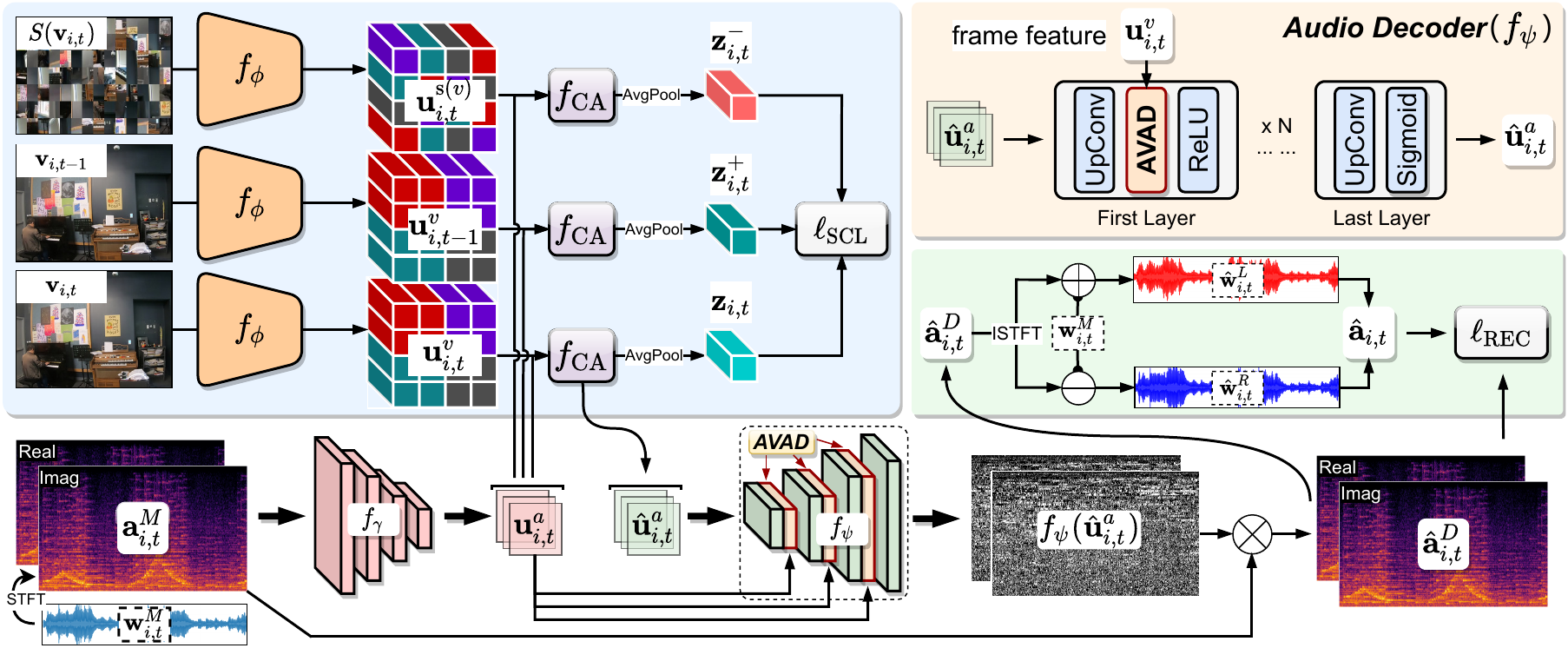} 
    \vspace{-10pt}
    \caption{
    Illustration of our \textbf{CCStereo} method during training. Given a pair of mono audio signals, $\mathbf{a}^{M}_{i,t}$, and a corresponding video frame, $\mathbf{v}_{i,t}$, as input, the objective is to predict the spectrogram of the difference audio, $\hat{\mathbf{a}}^{D}_{i,t}$, using a U-Net model~\cite{ronneberger2015u}. 
    The model comprises an image encoder network ($f_\phi$), an audio encoder network ($f_\gamma$), and an audio decoder network ($f_\psi$), which incorporates an Audio-Visual Adaptive De-normalisation (AVAD) layer to enhance feature adaptation.
    The overall training objective consists of two tasks: 1) accurately reconstructing $\hat{\mathbf{a}}^{D}_{i,t}$, and 2) a contrastive learning task aimed at learning discriminative representations concerning spatial changes.
    }
    \vspace{-15pt}
    \label{fig:framework}
    \Description[Overview of method CCStereo method]{Overview of method CCStereo method. The model comprises an encoder network, $f_\gamma$, and a decoder network, $f_\psi$, which incorporates an Audio-Visual Adaptive De-normalisation (AVAD) layer to enhance feature adaptation.}
\end{figure*}

\section{Related Works} 
\textbf{Binaural audio generation (BAG)} methods aim to create binaural audio from monaural recordings using visual information.
Mono2Binaural~\cite{gao20192}, the first binaural audio generation method, uses a U-Net~\cite{ronneberger2015u} to estimate the differential audio between the left and right channels by leveraging visual-spatial cues.
However, operations like tiling and concatenation at the bottleneck layer~\cite{zhou2020sep} and average pooling~\cite{gholamalinezhad2020pooling} can lead to overfitting~\cite{wang2020makes,cui2023deep} and loss of spatial details~\cite{zhao2024improved}, limiting the model’s ability to capture complex spatial relationships.
Enhancing the use of visual information in binaural audio generation has been a primary focus of recent research~\cite{zhang2021multi,li2021binaural,garg2021geometry,parida2022beyond,li2024cyclic}. Various methods are proposed to improve the model's understanding of semantic and spatial information. 
These methods can broadly be categorised into three major directions: 1) improving cross-modal feature interaction~\cite{zhang2021multi} via attention mechanism~\cite{vaswani2017attention} to better fuse the information between audio and visual modalities; 2) 
employing proxy learning tasks that help the model better understand the spatial correlation between the two modalities, such as discriminating the position of sound sources~\cite{liu2024visually} or identifying their locations~\cite{li2024cyclic}; and 3) introducing the geometry clue of the scene, such as depth information~\cite{parida2022beyond} and room impulse response~\cite{garg2021geometry} to leverage the 3D environment during model reasoning.
However, overfitting to the visual environment remains a challenge, potentially hindering the model's generalisation ability.
Additionally, prior studies~\cite{zhou2020sep, xu2021visually} have pointed out challenges like limited data availability and overfitting to visual environments. Efforts have been made to tackle these issues by using external monaural datasets~\cite{zhou2020sep} and reorganising benchmarks~\cite{xu2021visually} to enhance model robustness and generalisation evaluation. 
Despite their efficiency, these methods~\cite{zhou2020sep, xu2021visually} still heavily rely on cross-attention to guide the decoding process. 
The cross-attention mechanism is effective at capturing alignment relationships across modalities~\cite{baltruvsaitis2018multimodal}. However, in tasks such as text-to-image generation, it has been shown to result in coarse-grained controllability when using a reference image~\cite{ye2023ip}. We argue that a similar limitation exists in binaural audio generation: cross-attention alone lacks explicit control over the spatial characteristics of the generated audio, which may lead to sub-optimal performance.

\textbf{Conditional normalisation layers} have been studied in style transfer~\cite{nam2018batch} and conditional image synthesis~\cite{park2019semantic}. Unlike standard normalisation methods~\cite{nam2018batch} that rely on batch or instance statistics (e.g., mean and variance), conditional normalisation modulates these statistics through an affine transformation learned from external conditioning data~\cite{abdal2021styleflow}. 
In semantic image synthesis~\cite{park2019semantic,wang2022semantic, tan2021diverse, dong2017semantic} and style transfer~\cite{huang2017arbitrary, gatys2016image, fu2018style, karras2019style}, this modulation is typically conditioned on semantic segmentation maps~\cite{park2019semantic}, style features~\cite{karras2019style,huang2017arbitrary}, or text descriptions~\cite{zhang2023adding,ye2023ip}, enabling the preservation of semantic structure during decoding.
Inspired by these successes, we propose an audio-visual normalisation strategy that operates in tandem with cross-attention layers for the audio generation process, where visual context modulates the feature statistics to complement attention-based fusion, enabling finer spatial control and more precise spatial audio generation.

\textbf{Contrastive learning} has emerged as a powerful self-supervised learning framework that enables models to learn meaningful representations by distinguishing between similar and dissimilar pairs~\cite{chen2020simple,chen2021exploring,he2020momentum}. Contrastive learning has also shown promising performance in audio-visual learning methods~\cite{chen2021localizing,chen2024unraveling,chen2021localizing,hu2022mix,mo2022localizing,mo2022closer}, aligning augmented representations of the same instance as positives while separating those of different instances as negatives within a batch. Binaural audio generation can similarly benefit from self-supervised learning tasks by leveraging contrastive objectives to distinguish left and right information in both audio~\cite{li2021binaural} and visual~\cite{liu2024visually} modalities.
In our work, we propose a novel self-supervised contrastive learning approach~\cite{chen2020simple,chen2021exploring,he2020momentum} that mines a large number of negative samples from temporally adjacent frames and spatially shuffled visual features to simulate changes in object position. 
Hence, it helps address the challenge of accurately disentangling spatial cues from noisy or ambiguous visual contexts, which is critical for tasks such as binaural audio generation and spatial sound understanding.

\textbf{Test-time augmentation (TTA)} 
improves model performance by applying data augmentation at inference, creating multiple variations of the input and aggregating predictions. TTA is widely used in computer vision to enhance robustness without additional training~\cite{kimura2021understanding}. 
Studies have shown that TTA effectively improves prediction robustness~\cite{shanmugam2021better}, though it comes at the cost of significantly reduced inference speed. 
To handle moving sound sources and camera motion, previous binaural audio generation methods~\cite{gao20192,zhou2020sep,xu2021visually,li2021binaural,li2024cyclic,liu2024visually} often adopted a sliding window strategy with a small hop size (e.g., $0.05$ seconds), which leads to a large number of duplicated frames. 
We leverage this unique inference characteristic to integrate TTA into the process without incurring additional computational costs, thereby enhancing model performance.

\section{Method}
We denote an unlabelled video dataset as $\mathcal{D}=\{(\mathbf{w}_i, \mathbf{v}_i)\}_{i=1}^{|\mathcal{D}|}$, where $\mathbf{v}_i\in\mathcal{V}\subset \mathbb{R}^{T\times H \times W \times 3}$ is a set of $T$ RGB images with resolution $H \times W$, 
$\mathbf{w}_i\in\mathcal{W}\subset \mathbb{R}^{C \times T'}$ denotes the waveform data with $C\in\{L,R\}$ channels and total number of $T'$ samples.
Given monaural audio ($\mathbf{w}^{M}_{i} = \mathbf{w}^{L}_{i} + \mathbf{w}^{R}_{i})$, 
We apply the short-time Fourier transform (STFT)~\cite{griffin1984signal} on $\mathbf{w}^{M}_{i}$, resulting in $\mathbf{a}^M_i \in \mathcal{A} \subset \mathbb{C}^{T_s \times F}$, where $F$ is the number of frequency bins and $T_s$ denotes the number of time frames.
Here, $\mathcal{V}$, $\mathcal{W}$ and $\mathcal{A}$ denote the spaces of visual data, audio waveform data, and audio spectrogram data, respectively.
The model predicts the spectrogram of the target difference audio, defined as $\mathbf{a}^{D}_{i} = \text{STFT}(\mathbf{w}^{L}_{i} - \mathbf{w}^{R}_{i})$.


\subsection{Preliminaries}

During training, we randomly sample an audio segment and its corresponding frame start at time step $t\in\mathcal{T}$ from each video (i.e., $(\mathbf{a}^M_{i,t}, \mathbf{v}_{i,t})$) to form an input pair for the model. 
Our goal is to learn the parameters $\theta \in \Theta$ for the model $f_{\theta}:\mathcal{V} \times \mathcal{A} \to [-1, 1]^{F \times T_{s}}$, which comprises the image and audio encoder that extract features with $\mathbf{u}^{a}_{i,t}=f_\gamma(\mathbf{a}^M_{i,t})$ and $\mathbf{u}^{v}_{i,t}=f_\phi(\mathbf{v}_{i,t})$, respectively, where $\gamma,\phi \in \theta$, and $\mathbf{u}^{a}_{i,t},\mathbf{u}^{v}_{i,t} \in \mathcal{U}$, with $\mathcal{U}$ denoting a unified feature space. Our approach adopted a multi-head attention block~\cite{vaswani2017attention}, which estimates the co-occurrence of audio and visual data. We simply define the cross-attention process as $\hat{\mathbf{u}}^{a}_{i,t} = f_{\text{CA}}(\mathbf{u}^{a}_{i,t},\mathbf{u}^{v}_{i,t})$, where $\mathbf{u}^{a}_{i,t}$ represent the query and $\mathbf{u}^{v}_{i,t}$ is the key and value.
We decode the $\hat{\mathbf{u}}^{a}_{i,t}$ through an audio decoder $\hat{\mathbf{a}}^{D}_{i,t} = f_{\psi}(\hat{\mathbf{u}}^{a}_{i,t}) \cdot \mathbf{a}^{M}_{i,t}$
, where $\psi\in\theta$.
Similar to previous methods~\cite{gao20192,zhou2020sep,xu2021visually,li2021binaural,li2024cyclic,liu2024visually}, we use the MSE loss,
\begin{equation}
    \ell_{\text{MSE}}(\mathbf{a}^D_{i,t}, \hat{\mathbf{a}}^D_{i,t}) = \frac{1}{L} \sum (\mathbf{a}^D_{i,t} - \hat{\mathbf{a}}^D_{i,t})^2 ,
\end{equation}
to constrain the U-Net's prediction for difference audio generation.
However, we empirically observed that constraining only the predicted difference audio might be sub-optimal. 
While predicting the interaural difference can help avoid degenerate solutions (e.g., identical-channel outputs), it does not explicitly enforce accurate modelling of spatial cues such as interaural time difference (ITD) or phase offset. Using naive MSE loss may lead to blurred or unstable spectral predictions (see Fig.~\ref{fig:rec_loss_visual}), especially in high-frequency regions, resulting in unstable localisation or cancellation effects due to inaccurate ITD reconstruction. 
To avoid the aforementioned issues, we introduce a magnitude loss~\cite{richard2021neural} on the predicted difference audio:
\begin{equation}
    \ell_{\text{APM}}(\mathbf{a}^D_{i,t}, \hat{\mathbf{a}}^D_{i,t}) = \frac{1}{L} \sum \left\lvert \lvert \mathbf{a}^D_{i,t} \rvert - \lvert \hat{\mathbf{a}}^D_{i,t} \rvert \right\rvert .
\end{equation}
This loss encourages the model to match the spectral energy distribution of the ground truth
and guides the model towards reconstructing more accurate and structured frequency representations, particularly in high-frequency regions where phase variations are rapid and energy is sparse. Here, $L = T \times F$, and $\lvert \cdot \rvert$ denotes the modulus of a complex number.
Additionally, we further add a phase loss to directly supervise the predicted binaural spectrogram $\hat{\mathbf{a}}_{i,t}$ against the ground truth $\mathbf{a}_{i,t}$. This objective encourages better phase alignment between the two:
\begin{equation}
    \ell_{\text{PHS}}(\mathbf{a}_{i,t}, \hat{\mathbf{a}}_{i,t}) = \frac{1}{L} \sum \|\angle(\mathbf{a}_{i,t}) - \angle(\hat{\mathbf{a}}_{i,t})\|_2,
\end{equation}
where $\angle(\cdot)$ denotes the phase angle of a complex spectrogram.
We denote the overall reconstruction loss as $\ell_{\text{REC}} = \ell_{\text{MSE}}+ \zeta \ell_{\text{APM}}+\eta \ell_{\text{PHS}}$, where $\zeta$ and $\eta$ are hyper-parameters.

\begin{figure}[t]
    \centering
    \includegraphics[width=.88\linewidth]{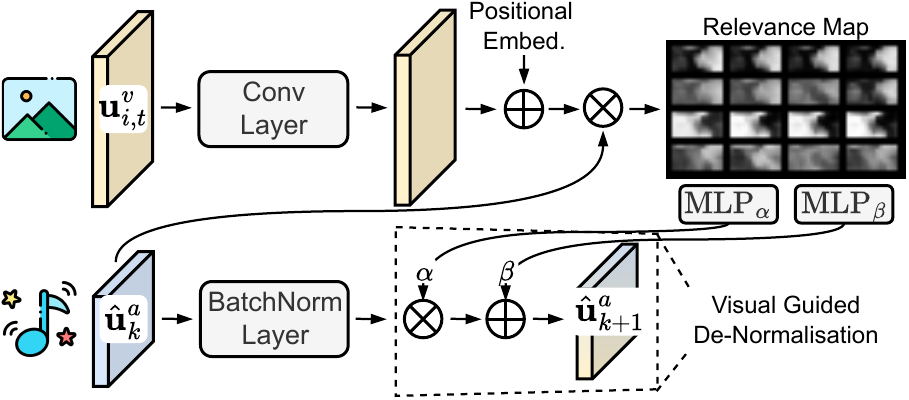} 
    \vspace{-10pt}
    \caption{
    Illustration of our \textbf{AVAD} layer. Unlike previous methods~\cite{gao20192,zhou2020sep,xu2021visually,li2024cyclic} that normalise over the batch, we introduce a de-normalisation process to refine spatial infusion during decoding. A relevance map computed between audio and visual features modulates the mean and variance, ensuring more precise spatial conditioning in the generated audio. 
    Each relevance map encodes the influence of a local pixel region relative to the audio feature.
    }
    \vspace{-15pt}
    \label{fig:avad}
    \Description[Illustration of our \textbf{AVAD} module during training.]{Illustration of our \textbf{AVAD} module during training.}
\end{figure}

\subsection{Audio-Visual Adaptive De-normalisation}

Unlike previous methods~\cite{gao20192,zhou2020sep} that rely solely on cross-attention or feature concatenation to fuse spatial information from the visual modality into audio, our audio-visual adaptive de-normalisation (AVAD) module aims to control the audio decoding process by modulating the statistics of local feature representations.
As illustrated in Fig.~\ref{fig:framework}, AVAD is integrated into the intermediate layers of the U-Net decoder $f_{\psi}$ by replacing standard batch normalisation layers with a visually informed de-normalisation module. This design allows the network to effectively incorporate both spatial and semantic cues from the visual modality into the decoding process.
For simplicity, we omit the subscripts $i$ and $t$ in the following.

The detailed module design is depicted in Fig.~\ref{fig:avad}. We first pass the audio feature map $\hat{\mathbf{u}}^a_k$ through a batch normalisation layer ($\text{BN}$) at the $k$-th layer, and then scale and shift the normalised feature using the estimated $\alpha$ and $\beta$ via 
\begin{equation}   
    \tilde{\mathbf{u}}^a_{k+1} = (1 + \alpha) \cdot \text{BN}(\tilde{\mathbf{u}}^a_k) + \beta .
\end{equation}
To dynamically adapt the normalisation parameters based on cross-modal context, we propose to compute the scale ($\alpha$) and shift ($\beta$) tensors using an audio-visual relevance map. Specifically, we first calculate a relevance map $\mathbf{c}_k = \tilde{\mathbf{u}}^a_k \cdot (\text{Conv}(\mathbf{u}^v)+\mathbf{p}_v)$, which captures the interaction between audio features and visual guidance at the layer $k$, where $\mathbf{p}_v$ denotes the positional embedding.
We then feed this relevance map into a shared MLP, followed by two modality-specific branches to estimate the affine parameters:
\begin{equation}
\begin{aligned}
    \alpha & = \text{MLP}_{\alpha}(\text{MLP}_{\text{share}}(\mathbf{c}_k)) \\
    \beta  & = \text{MLP}_{\beta}(\text{MLP}_{\text{share}}(\mathbf{c}_k))
\end{aligned}
\end{equation}

\subsection{Spatial-aware Contrastive Learning (SCL)}
The capability to learn discriminative feature presentation is crucial for audio-visual systems. One limitation of prior \revise{self-supervised methods for} binaural audio generation is their exclusive focus on proxy tasks within the audio domain (e.g., classifying whether the audio channels are flipped).
\revise{This narrow focus not only under-utilises visual positional information but also impedes the learning of a joint audio-visual representation.}
We argue that two requirements must be satisfied to achieve effective contrastive learning: 1) spatial awareness in the learned joint representation and 2) inclusion of a diverse set of examples. 
Unfortunately, previous audio-visual contrastive learning methods~\cite{chen2021localizing,chen2024unraveling} may not be suitable for the current task, as they generally failed to satisfy these two requirements. 

Motivated by the observation that spatial misalignment between audio and visual features disrupts the perception of coherent cross-modal correspondence, we design a shuffle-based contrastive strategy that introduces spatial perturbations to generate informative negatives and promote spatially grounded learning.
Since the BAG problem cannot access video-level labels, we adopt a classic instance discrimination pipeline (e.g., SimCLR~\cite{chen2020simple}), where each audio-visual pair is treated as an independent contrastive class. For a randomly sampled minibatch of $N$ examples, we perform the contrastive prediction task on pairs of positive and negative pairs derived from the minibatch. 
We define the anchor set $\mathcal{E}$, positive set $\mathcal{P}$ and negative set $\mathcal{N}$ as follows:
\begin{equation}
\scalebox{1.0}{$
\begin{aligned}
    \mathcal{E} & = \left\{ \mathbf{z}_{i,t} \mid \mathbf{z}_{i,t} = p\left(f_{\text{CA}}(\mathbf{u}^{a}_{i,t}, f_\phi(\mathbf{v}_{i,t}))\right), i \in \mathcal{D}, t \in \mathcal{T} \right\}, \\
    \mathcal{P} & = \left\{ \mathbf{z}^{+}_{i,t} \mid \mathbf{z}^{+}_{i,t} = p\left(f_{\text{CA}}(\mathbf{u}^{a}_{i,t}, f_\phi(\mathbf{v}_{i,t-1}))\right), i \in \mathcal{D}, t \in \mathcal{T} \right\}, \\
    \mathcal{N} & = \left\{ \mathbf{z}^{-}_{i,t} \mid \mathbf{z}^{-}_{i,t} = p\left(f_{\text{CA}}(\mathbf{u}^{a}_{i,t}, f_\phi(S(\mathbf{v}_{i,t})))\right), i \in \mathcal{D}, t \in \mathcal{T} \right\},
\end{aligned}
$}
\label{eq:sets}
\end{equation}
where $p(\cdot)$ is the 2D average pooling and $S(\cdot)$ represent a shuffle process over the spatial dimension $H$ and $W$ of $\mathbf{v}_{i,t}$. Adopting the InfoNCE~\cite{chen2020simple}, we define the objective function as follows:
\begin{equation}
\scalebox{0.98}{$
\begin{aligned}
\ell_{\text{SCL}}(\mathbf{z}_j) = - 
\log \frac{
    \exp\left(\mathbf{z}_j\cdot \mathbf{z}_j^+ / \tau\right)
}{
    \exp\left(\mathbf{z}_j \cdot \mathbf{z}_j^+ / \tau\right)
    + \sum_{\mathbf{z}_n^- \in \mathcal{N}} \exp\left(\mathbf{z}_j\cdot \mathbf{z}_n^- / \tau\right)
},
\end{aligned}
$}
\label{eq:loss_scl}
\end{equation}
where $\mathbf{z}_j \in \mathcal{E}$ is an anchor feature, $\mathbf{z}_j^+ \in \mathcal{P}$ is its corresponding positive pair, $\mathbf{z}_n^- \in \mathcal{N}$ are the negative features, and $\tau = 0.1$ is the temperature hyper-parameter.

\begin{figure}
    \centering
    \includegraphics[width=.9\linewidth]{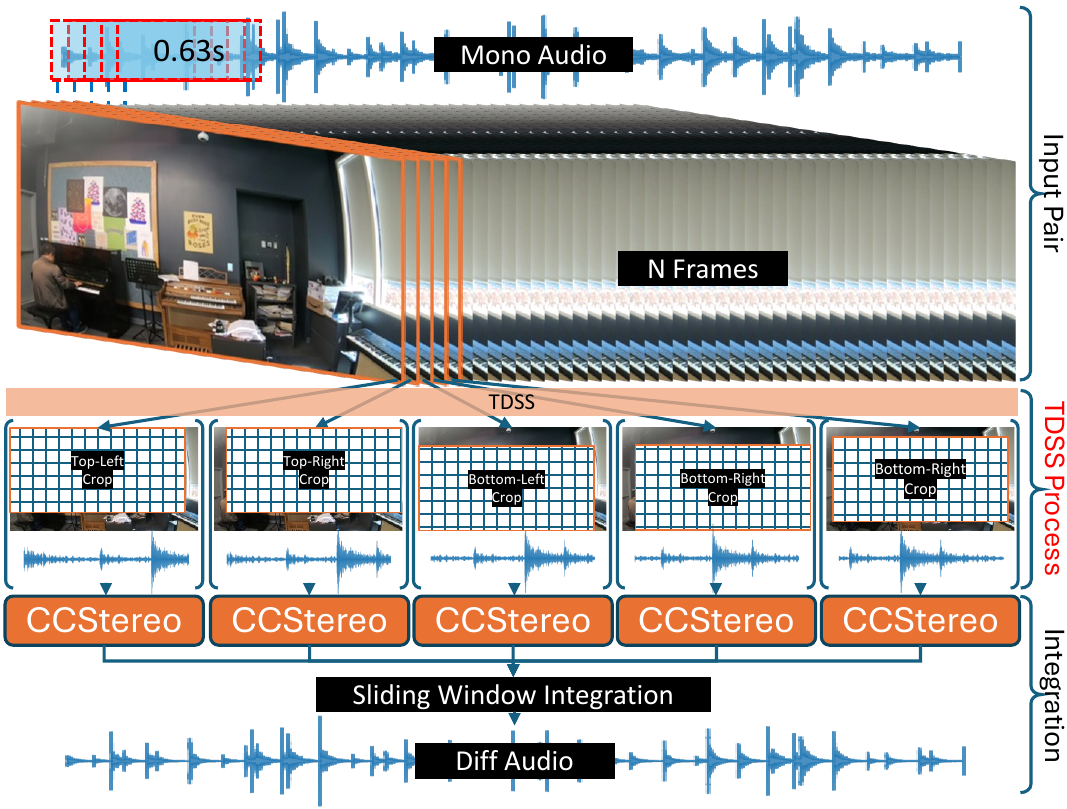}
    \vspace{-10pt}
    \caption{Overview of the model inference procedure.}
    \label{fig:infer}
    \Description[Overview of the model inference procedure.]{Overview of the model inference procedure.}
    \vspace{-15pt}
\end{figure}

\subsection{Training \& Testing}

\textbf{Overall Training.}
The overall training objective is $\ell = \ell_{\text{REC}} + \lambda\ell_{\text{SCL}}$, where $\lambda$ is a hyper-parameter.

\textbf{Test-time Dynamic Scene Simulation (TDSS).}
During inference, we firstly estimate the left and right complex spectrograms through $\mathbf{\hat{a}}^{L}_i = (\mathbf{a}^{M}_{i} + \mathbf{\hat{a}}^{D}_i) / 2$ and $\mathbf{\hat{a}}^{R}_i = (\mathbf{a}^{M}_{i} - \mathbf{\hat{a}}^{D}_i) / 2$. Then, we use inverse STFT (ISTFT)~\cite{griffin1984signal} to recover the audio signal from both channels and concatenate them together to form the final binaural waveform prediction $\mathbf{\hat{w}}_i = \text{Concat}[ISTFT(\mathbf{\hat{a}}^{L}_i), ISTFT(\mathbf{\hat{a}}^{R}_i)]$. We use a sliding window of 0.63 seconds and a hop size of 0.1 seconds to binauralise 10-second audio clips, following an approach similar to that of the baseline methods~\cite{gao20192}. 
While this process improves binaural audio generation by focusing on smaller audio segments, it introduces significant computational redundancy. Motivated by the small visual differences in 10 fps music videos, we design TDSS to leverage this redundancy for better performance and robustness.

As depicted in Fig.~\ref{fig:infer} and Alg.~\ref{alg:tdss}, instead of directly resizing every video frame to $448\times224$~\cite{gao20192,zhou2020sep,xu2021visually}, we first resize each frame to $480\times240$ and then crop a $448\times224$ window from one of the five regions  $[\text{top-left}, \text{top-right}, \text{bottom-left}, \text{bottom-right}, \text{centre}]$ based on the current frame index (i.e., ``$i$ \% 5'', where $i$ is the frame index). For example, if the first two audio segments are paired with the \nth{5} and \nth{6} frames, we crop the top-left corner of the \nth{5} frame and the top-right corner of the \nth{6} frame, respectively.
Please refer to the \textit{Supplementary Material} for additional details on sliding window integration.

\begin{algorithm}[t!]
\caption{Test-time Dynamic Scene Simulation (TDSS)}
\label{alg:tdss}
\definecolor{codeblue}{rgb}{0.25,0.5,0.5}
\definecolor{codekw}{rgb}{0.85, 0.18, 0.50}
\definecolor{ao}{rgb}{0.0, 0.5, 0.0}
\begin{algorithmic}[1]
    \Function{$\func{load\_frame}$}{$f, i$}
    \State \textcolor{ao}{\# $f$: current frame path}
    \State \textcolor{ao}{\# $i$: current frame index}
    \State \textcolor{ao}{\# frames are fixed at $448 \times 224$ resolution.}
    \State \revise{$\mathbf{v}_{i,t}$} = Image.open($f$)
    \State $w$, $h$ = image.size
    \State $w_\mathsf{crop}$,  $h_\mathsf{crop}$ = 448, 224
    \State $\mathsf{points}$ = [
    \State \hspace{1em}$(0, 0)$, \Comment{Top-Left}
    \State \hspace{1em}$(w - w_\mathsf{crop}, 0)$, \Comment{Top-Right}
    \State \hspace{1em}$(0, h - h_\mathsf{crop})$, \Comment{Bottom-Left}
    \State \hspace{1em}$(w - w_\mathsf{crop}, h - h_\mathsf{crop})$, \Comment{Bottom-Right}
    \State \hspace{1em}$((w - w_\mathsf{crop})\,//\,2, (h - h_\mathsf{crop})\,//\,2)$ \Comment{Center}
    \State ]
    \State $\func{point\_idx} = i\,\%\,\func{len}(\func{points})$
    \State $\func{point\_idx} = \max(0,\ \min(\func{point\_idx},\ \func{len}(\func{points}) - 1))$
    \State $\func{left},\ \func{upper} = \func{points}[\func{point\_idx}]$
    \State \Return \small{$\func{\mathbf{v}_{i,t}}.\func{crop}((\func{left},\ \func{upper},\ \func{left} + w_\mathsf{crop},\ \func{upper} + h_\mathsf{crop}))$}
    \EndFunction
\end{algorithmic}
\end{algorithm}

\section{Experiments}
\subsection{Evaluation Protocols}
\noindent \textbf{Datasets.}
We adopt three widely used music video datasets, FAIR-Play~\cite{gao20192}, MUSIC-Stereo~\cite{zhao2018sound,xu2021visually} and YT-MUSIC~\cite{morgado2018self,gao20192}, for the model evaluation process. 
The \textbf{FAIR-Play}~\cite{gao20192} dataset contains 1,871 10-second clips of videos recorded in a music room, with a total playtime of 5.2 hours. The videos were recorded using a professional binaural microphone, preserving high-quality binaural audio. The FAIR-Play dataset has two commonly used train/validation/test split setups. The first is the 10-split setup~\cite{gao20192}, which randomly divides the videos into subsets. The second is the 5-split setup~\cite{xu2021visually}, designed to evaluate the model's true generalisation ability by reducing scene overlap between training and testing, providing a more challenging evaluation setting. The videos are extracted to frames at 10 fps~\cite{gao20192,zhou2020sep}. 

We also evaluate our approach on the \textbf{MUSIC-Stereo} dataset~\cite{xu2021visually}, which is based on the MUSIC dataset~\cite{zhao2018sound} containing 21 types of musical instruments, featuring both solo and duet performances. We follow previous works~\cite{zhou2020sep,xu2021visually,parida2022beyond} by filtering out non-binaural cases using a threshold of 0.001 for the sum of left-right channel differences. We obtained 1,047 unique videos with binaural audio. We then divided the videos into 80-10-10 for training, validation, and testing. Following previous works~\cite{xu2021visually,parida2022beyond}, we split the videos into 10-second clips and finally arrived at 20,351 clips, which is 10$\times$ larger than the FAIR-Play dataset. 

\begin{table}[t]
    \centering
    \def\arraystretch{1.1}
    \caption{Comparison with existing approaches on \textbf{FAIR-Play (10-splits)}~\cite{gao20192,xu2021visually}. Where \ding{72} indicates the model uses extra data from MUSIC21-Solo~\cite{liu2024visually} dataset. Best results are shown in \textbf{bold}, and the $2^{nd}$ best are \underline{underlined}.}
    \vspace{-10pt}
    \resizebox{.98\linewidth}{!}{
    \begin{tabular}{!{\vrule width 1.1pt}l!{\vrule width 1.1pt}cccc!{\vrule width 1.1pt}}
    \specialrule{1.1pt}{0pt}{0pt}
    \multirow{2}{*}{Methods} & \multicolumn{4}{c!{\vrule width 1.1pt}}{FAIR-Play (10-splits)~\cite{gao20192,xu2021visually}} \\
    \cline{2-5}
    ~ & $\text{STFT} \downarrow$ & $\text{ENV} \downarrow$ & $\text{WAV} \downarrow$ & $\text{SNR} \uparrow $ \\
    \specialrule{1.1pt}{0pt}{0pt}
    Mono2Binaural~\cite{gao20192} & 0.959 & 0.141 & 6.496 & 6.232 \\ 
    APNet~\cite{zhou2020sep} & 0.889 & 0.136 & 5.758 & 6.972 \\ 
    Sep-stereo~\cite{zhou2020sep} \ding{72} & 0.879 & 0.135 & 6.526 & 6.422 \\ 
    Main Net.~\cite{zhang2021multi} & 0.867 & 0.135 & 5.750 & 6.985 \\ 
    Complete Net.~\cite{zhang2021multi} & 0.856 & 0.134 & 5.787 & 6.959 \\ 
    SAGM~\cite{li2024cross} & 0.851 & 0.134 & \underline{5.684} & \underline{7.044} \\ 
    CMC~\cite{liu2024visually} & \underline{0.849} & \underline{0.133} & - & - \\ 
    \hline
    \rowcolor{LightCyan}\textbf{CCStereo}  & \textbf{0.823} & \textbf{0.132} & \textbf{5.502} & \textbf{7.144} \\ 
    \specialrule{1.1pt}{0pt}{0pt}
    \end{tabular}
    }
    \label{tab:fairplay_10s}
    \vspace{-10pt}
 \end{table}
\begin{table}[t!]
    \centering
    \def\arraystretch{1.1}
    \caption{
    Comparison with existing approaches on \textbf{FAIR-Play (5-splits)}~\cite{gao20192,xu2021visually}. Where \ding{72} denotes that the model uses additional data from the MUSIC21-Solo~\cite{liu2024visually} dataset, and the results in \colorbox{gray!25}{gray} indicates a reproduced implementation of the method.
    Best results are shown in \textbf{bold}, and the $2^{nd}$ best are \underline{underlined}.}
    \vspace{-10pt}
    \resizebox{0.98\linewidth}{!}{
    \begin{tabular}{!{\vrule width 1.1pt}l!{\vrule width 1.1pt}ccccc!{\vrule width 1.1pt}}
    \specialrule{1.1pt}{0pt}{0pt}
    \multirow{2}{*}{Methods} & \multicolumn{5}{c!{\vrule width 1.1pt}}{FAIR-Play (5-splits)~\cite{gao20192,xu2021visually}} \\
    \cline{2-6}
    ~ & $\text{STFT} \downarrow$ & $\text{ENV} \downarrow$ & $\text{Mag} \downarrow$ & $ \text{Phs} \downarrow$ & $\text{SNR} \uparrow $ \\ 
    \specialrule{1.1pt}{0pt}{0pt}
    Mono-Mono~\cite{xu2021visually} & 1.024 & 0.145 & 2.049 & 1.571 & 4.968  \\ 
    Mono2Binaural~\cite{gao20192, xu2021visually} & 0.917 & \underline{0.137} & 1.835 & 1.504 & 5.203 \\ 
    PseudoBinaural~\cite{xu2021visually} & 0.944 & 0.139 & 1.901 & 1.522 & 5.124 \\ 
    Sep-Stereo~\cite{zhou2020sep} \ding{72} & \underline{0.906} & \textbf{0.136} & \underline{1.811} & 1.495 & 5.221 \\ 
    CMC~\cite{liu2024visually} & \cellcolor{gray!25} 0.912 & \cellcolor{gray!25} 0.141 & \cellcolor{gray!25} 1.824 & \cellcolor{gray!25} 1.502 & \cellcolor{gray!25} 6.238 \\
    BeyondM2B~\cite{parida2022beyond} & 
    0.909 & 0.139 & 1.819 & \underline{1.479} & \underline{6.397} \\ 
    \hline
    \rowcolor{LightCyan}\textbf{CCStereo} & \textbf{0.883} & \underline{0.137} & \textbf{1.766} & \textbf{1.454} & \textbf{6.475} \\
    \specialrule{1.1pt}{0pt}{0pt}
    \end{tabular}
    }
    \label{tab:fairplay_5s}
    \vspace{-13pt}
 \end{table}

We additionally evaluate our method on the \textbf{YT-MUSIC} dataset \cite{morgado2018self}, which consists of 360$\degree$ YouTube videos in the ambisonic format, featuring three types of video projections: Equi-Angular Cubemap (EAC), Equirectangular (EQR), and Equal-Area (ER). We observed that some projection format labels in the dataset are incorrect~\footnote{Also noted in \url{https://github.com/pedro-morgado/spatialaudiogen/issues/13}}. To address this, we manually reclassified each video to ensure accurate labeling. Following prior works~\cite{gao20192,xu2021visually}, we use the official train-test split and preprocess the videos into 10-second clips, resulting in 8,681 training clips, 2,909 validation clips, and 2,909 testing clips. We follow previous works~\cite{gao20192,xu2021visually} in decoding ambisonic audio into binaural format.
As the MUSIC-Stereo~\cite{zhao2018sound} and YT-MUSIC~\cite{morgado2018self} datasets are YouTube-based datasets, the total number of available samples may fluctuate. The videos are extracted to frames at 10 fps~\cite{xu2021visually}.

\begin{table}[t]
    \centering
    \def\arraystretch{1.1}
    \caption{
    Comparison with existing approaches on \textbf{MUSIC-Stereo} dataset~\cite{zhao2018sound,xu2021visually}. Where \ding{72} denotes that the model uses additional data from the MUSIC21-Solo~\cite{liu2024visually} dataset.
    Best results are shown in \textbf{bold}, and the $2^{nd}$ best are \underline{underlined}.}
    \vspace{-10pt}
    \resizebox{0.98\linewidth}{!}{
    \begin{tabular}{!{\vrule width 1.1pt}l!{\vrule width 1.1pt}ccccc!{\vrule width 1.1pt}}
    \specialrule{1.1pt}{0pt}{0pt}
    %
    %
    \multirow{2}{*}{Methods} & \multicolumn{5}{c!{\vrule width 1.1pt}}{MUSIC-Stereo~\cite{zhao2018sound,xu2021visually}} \\
    \cline{2-6}
    ~ & $\text{STFT} \downarrow$ & $\text{ENV} \downarrow$ & $\text{Mag} \downarrow$ & $ \text{Phs} \downarrow$ & $\text{SNR} \uparrow $ \\
    \specialrule{1.1pt}{0pt}{0pt}
    Mono-Mono~\cite{xu2021visually}  & 1.014 & 0.144 & 2.027 & 1.568 & 7.858 \\ 
    Mono2Binaural~\cite{gao20192, xu2021visually}  & 0.942 & 0.138 & 1.885 & 1.550 & 8.255 \\ 
    PseudoBinaural~\cite{xu2021visually}  & 0.943 & 0.139 & 1.886 & 1.562 & 8.198 \\ 
    Sep-Stereo~\cite{zhou2020sep} \ding{72} & 0.929 & 0.135 & 1.803 & 1.544 & 8.306 \\ 
    CMC~\cite{liu2024visually} & 0.759 & 0.113 & 1.518 & \textbf{1.502} & - \\
    BeyondM2B~\cite{parida2022beyond} & \underline{0.670} & \underline{0.108} & \underline{1.340} & \underline{1.538} & \underline{10.754} \\ 
    \hline
    \rowcolor{LightCyan}\textbf{CCStereo} & \textbf{0.624} & \textbf{0.097} & \textbf{1.248} & 1.578 & \textbf{12.985} \\
    \specialrule{1.1pt}{0pt}{0pt}
    \end{tabular}
    }
    \label{tab:music_stereo}
    \vspace{-12pt}
 \end{table}
 \begin{table}[t]
    \centering
    \def\arraystretch{1.1}
    \caption{Comparison with existing approaches on \textbf{YT-MUSIC}~\cite{morgado2018self}. Where \ding{72} indicates the model uses extra data from MUSIC21-Solo~\cite{liu2024visually} dataset. Best results are shown in \textbf{bold}, and the $2^{nd}$ best are \underline{underlined}. 
    }
    \vspace{-10pt}
    \resizebox{0.98\linewidth}{!}{
    \begin{tabular}{!{\vrule width 1.1pt}l!{\vrule width 1.1pt}ccccc!{\vrule width 1.1pt}}
    \specialrule{1.1pt}{0pt}{0pt}
    \multirow{2}{*}{Methods} & \multicolumn{5}{c!{\vrule width 1.1pt}}{YT-MUSIC~\cite{morgado2018self}
    } \\
    \cline{2-6}
    ~ & $\text{STFT} \downarrow$ & $\text{ENV} \downarrow$ & $\text{Mag} \downarrow$ & $ \text{Phs} \downarrow$ & $\text{SNR} \uparrow $ \\
    \specialrule{1.1pt}{0pt}{0pt}
    Mono2Binaural & 0.501 & 0.110 & 1.002 & 0.963 & 6.712  \\ 
    PseudoBinaural~\cite{xu2021visually}  & 0.489 & 0.109 & 0.979 & 0.922 & 7.610  \\ 
    Sep-Stereo~\cite{zhou2020sep} \ding{72} & \underline{0.466} & \underline{0.106} & \underline{0.933} & \underline{0.917} & \underline{7.844}  \\ 
    %
    \hline
    \rowcolor{LightCyan}\textbf{CCStereo} & \textbf{0.432} & \textbf{0.102} & \textbf{0.865} & \textbf{0.854} & \textbf{8.245} \\
    \specialrule{1.1pt}{0pt}{0pt}
    \end{tabular}
    }
    \vspace{-15pt}
    \label{tab:yt_music}
 \end{table}

\noindent \textbf{Evaluation Metrics.}
We follow the previous methods~\cite{zhou2020sep,li2024cross,xu2021visually,parida2022beyond} to report the \textit{STFT L2 distance ($\text{STFT}$)}, \textit{Magnitude distance ($\text{Mag}$)} and \textit{Difference Phase Distance ($\text{Phs}$)} on the time-frequency domain; and \textit{waveform L2 distance ($\text{WAV}$)}, \textit{envelope distance ($\text{ENV}$)} and \textit{Signal-to-Noise Ratio ($\text{SNR}$)} on time domain to assess the fidelity and quality of generated binaural audios. 
Please note that on FAIR-Play (10-splits)~\cite{gao20192}, we adopt $\text{WAV}$ in place of $\text{Mag}$ and $\text{Phs}$, following previous benchmarks~\cite{gao20192,zhou2020sep,li2021binaural,liu2024visually}, to enable a consistent comparison.

\begin{table*}[t]
\centering
\caption{Ablation study of the model components on FAIR-Play (5-splits)~\cite{gao20192} split 2 and MUSIC-Stereo~\cite{gao20192,xu2021visually}. 
}
\vspace{-10pt}
\def\arraystretch{1.1}
\resizebox{0.9\linewidth}{!}{%
\begin{tabular}{
!{\vrule width 1.2pt}P{30pt}|P{30pt}|P{30pt}|P{30pt}|P{30pt}!{\vrule width 1.2pt}ccccc!{\vrule width 1.2pt}ccccc!{\vrule width 1.2pt}
} 
\specialrule{1.2pt}{0pt}{0pt}
\multicolumn{5}{!{\vrule width 1.2pt}c!{\vrule width 1.2pt}}{Method} & \multicolumn{5}{c!{\vrule width 1.2pt}}{FAIR-Play (5-splits)~\cite{gao20192,xu2021visually}} & \multicolumn{5}{c!{\vrule width 1.2pt}}{MUSIC-Stereo~\cite{gao20192,xu2021visually}}  \\ \hline
Baseline & TDSS & AVAD & $\ell_{\text{REC}}$   & $\ell_{\text{SCL}}$ 
& $\text{STFT} \downarrow$ & $\text{ENV} \downarrow$ & $\text{Mag} \downarrow$ & $ \text{Phs} \downarrow$ & $\text{SNR} \uparrow $ 
& $\text{STFT} \downarrow$ & $\text{ENV} \downarrow$ & $\text{Mag} \downarrow$ & $ \text{Phs} \downarrow$ & $\text{SNR} \uparrow $
\\ \hline
%
\redtick &  &  &  &  & 0.941 & 0.145 & 1.881 & 1.525 & 6.043 
                     & 0.653 & 0.104 & 1.306 & 1.557 & 11.972
\\ 
\redtick & \redtick &  &  &  & 0.917 & 0.142 & 1.834 & 1.493 & 6.179 
                             & 0.647 & 0.098 & 1.294 & 1.560 & 11.669
\\ 
\redtick & \redtick & \redtick &  &  & 0.908 & 0.140 & 1.815 & 1.486 & 6.254 
                                     & 0.638 & 0.102 & 1.268 & 1.586 & 12.698
\\ 
\redtick & \redtick & \redtick & \redtick &  & 0.891 & 0.139 & 1.783 & 1.453 & 6.371 
                                             & 0.630 & 0.098 & 1.260 & 1.580 & 12.960
\\ 
\redtick & \redtick & \redtick & \redtick & \redtick & \textbf{0.885} & \textbf{0.138} & \textbf{1.771} & \textbf{1.451} & \textbf{6.457} 
                                                     & \textbf{0.624} & \textbf{0.097} & \textbf{1.248} & \textbf{1.578} & \textbf{12.985}
\\ 
\specialrule{1.2pt}{0pt}{0pt}
\end{tabular}
\label{tab:ablation_components}
}
\vspace{-10pt}
\end{table*}

\begin{figure*}[t]
    \centering
    \begin{subfigure}[t]{0.45\linewidth}
        \centering
        \includegraphics[width=\textwidth]{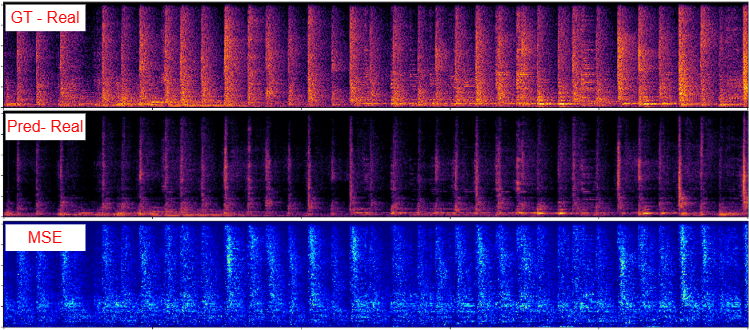}
        \caption{With MSE Loss}
        \vspace{-12pt}
        \label{fig:ablation_without_rec}
    \end{subfigure}%
    \begin{subfigure}[t]{0.45\linewidth}
        \centering
        \includegraphics[width=\textwidth]{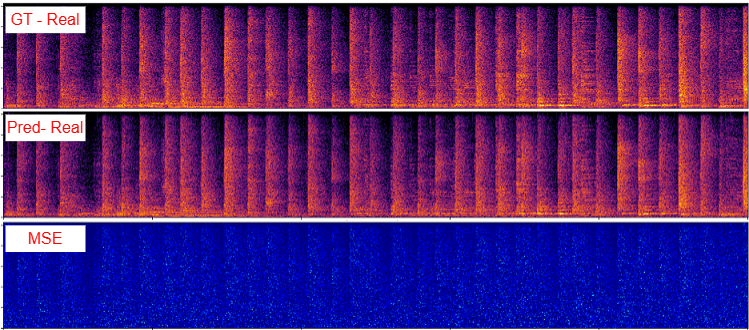}
        \caption{With REC Loss}
        \vspace{-12pt}
        \label{fig:ablation_with_rec}
    \end{subfigure}
    \caption{Qualitative comparison of predicted real spectrograms under different loss settings.}
    \Description[]{}
    \label{fig:rec_loss_visual}
    \vspace{-12pt}
\end{figure*}

\subsection{Implementation Details}
We follow previous methods~\cite{zhou2020sep,li2024cross,xu2021visually,parida2022beyond}
to fix the audio sampling rate to 16 kHz and normalise each segment’s RMS level to a constant value. We adopted a widely used audio preprocessing protocol by applying the STFT with a Hann window of 25 ms, a hop length of 10 ms, and an FFT size of 512. During training, we randomly sample 0.63-second audio segments from each 10-second clip, along with the corresponding central visual frame. The selected frame is resized to 480×240, then randomly cropped to 448×224. We also apply colour and intensity jittering as data augmentation, following~\cite{gao20192}.
We use a convolutional U-Net architecture~\cite{gao20192} for the audio backbone and a ResNet~\cite{he2016deep} (pre-trained on ImageNet~\cite{deng2009imagenet}) for the image backbone. 
The networks are trained using the Adam optimiser with a learning rate of 5e-5 for the image backbone and 5e-4 for the audio backbone, using a batch size of 128. 
We empirically set $\lambda$ to 0.1, $\zeta$ to 0.005 and $\eta$ to 1.0.

\begin{figure}[t]
    \centering
    \includegraphics[width=0.9\linewidth]{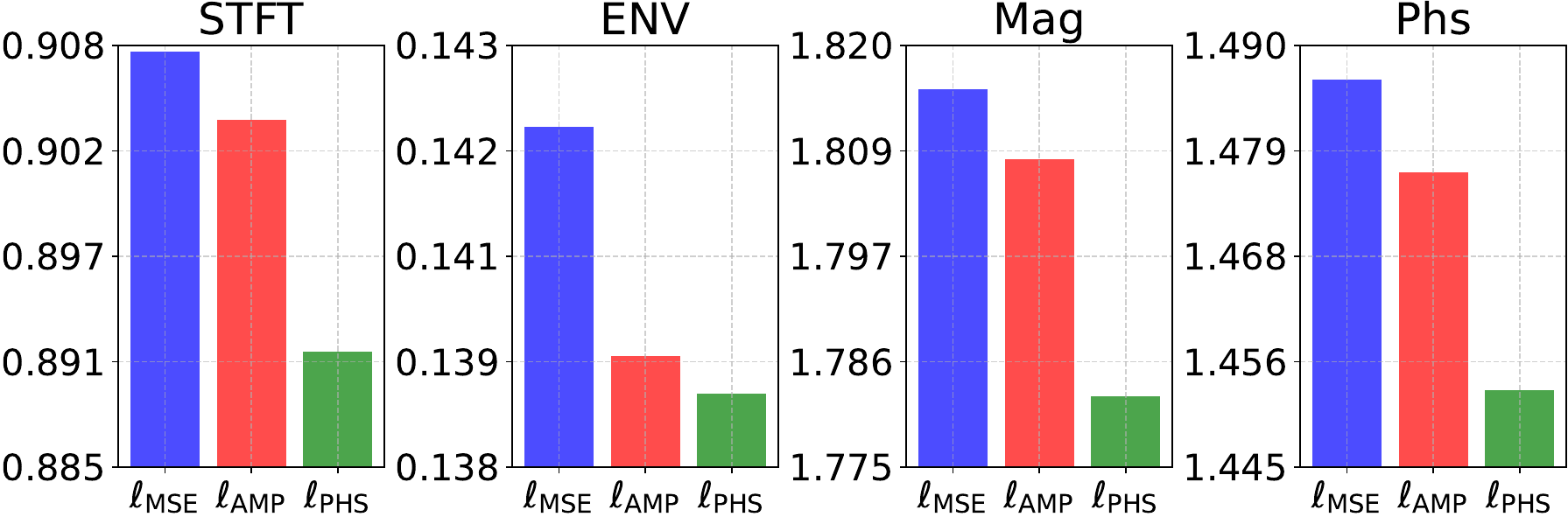}
    \vspace{-10pt}
    \caption{
    Ablation study on the $\ell_{\text{REC}}$ loss for the FAIR-Play dataset (5-splits) ~\cite{gao20192,xu2021visually}. The evaluation begins with $\ell_{\text{MSE}}$ (\textcolor{blue}{blue}), and sequentially adds $\ell_{\text{AMP}}$ (\textcolor{red}{red}) and $\ell_{\text{PHS}}$ (\textcolor{ForestGreen}{green}).
    }
    \vspace{-15pt}
    \label{fig:ablation_rec_loss}
    \Description{Ablation study on the $\ell_{\text{REC}}$ loss for the FAIR-Play dataset.}
\end{figure}

\begin{figure*}[t]
    \centering
    \includegraphics[width=.88\linewidth, height=5cm]{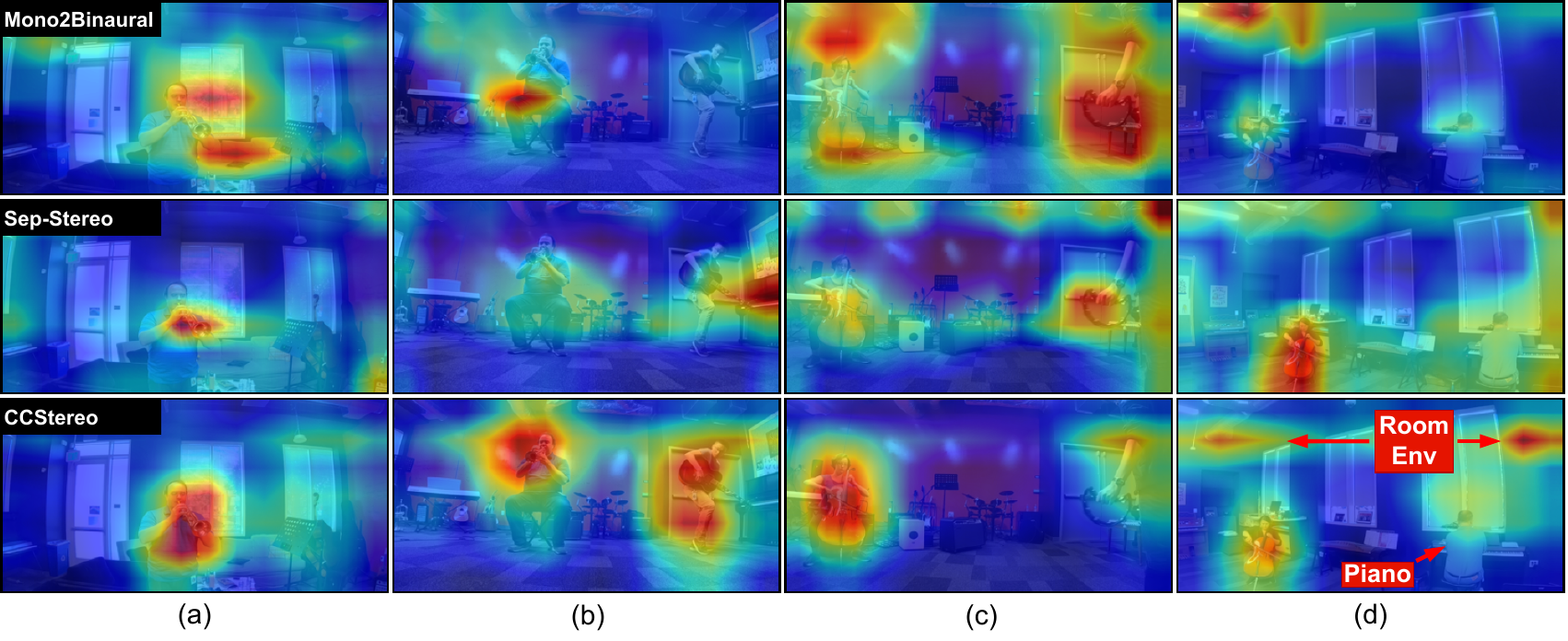} 
    \vspace{-10pt}
    \caption{
    Visual comparison of visual feature activation between Mono2Binaural~\cite{gao20192} (\nth{1} row), Sep-Stereo~\cite{zhou2020sep} (\nth{2} row) and CCStereo (\nth{3} row) on the FAIR-Play dataset~\cite{gao20192}.
    }
    \label{fig:attn_visual}
    \Description{Visual comparison of visual feature activation map.}
    \vspace{-10pt}
\end{figure*}

\begin{figure}[t!]
    \centering
    \begin{subfigure}[b]{\linewidth}
        \centering
        \includegraphics[width=0.92\textwidth, height=1.885cm]{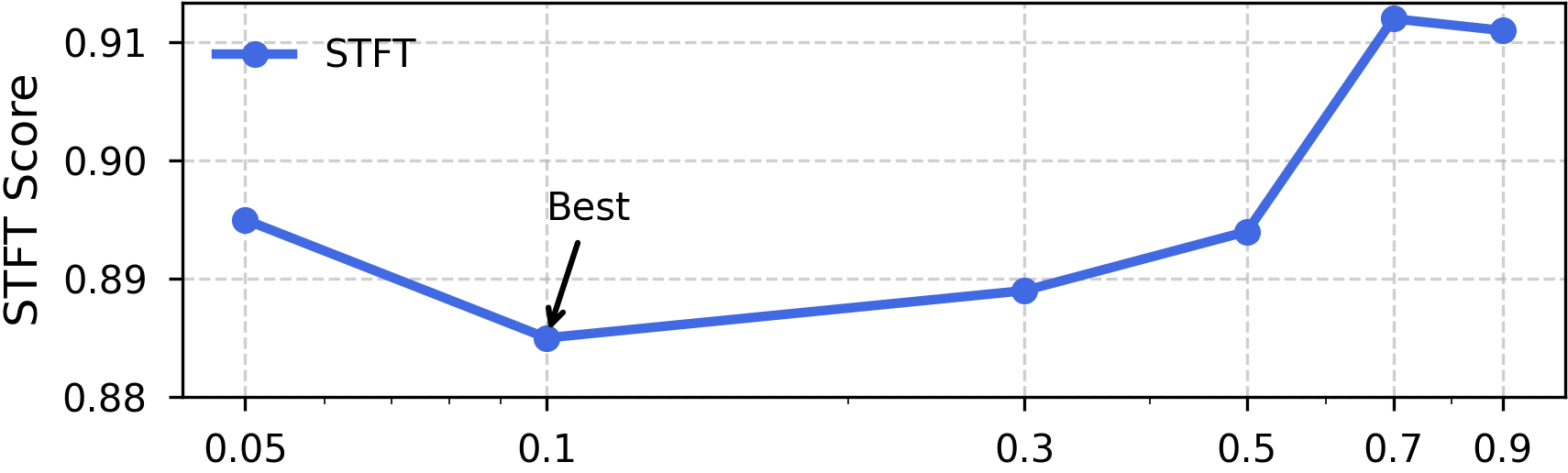}
        \vspace{-6pt}
        \caption{Hyper-parameter $\lambda$}
        \label{fig:ablation_lambda}
    \end{subfigure}
    \begin{subfigure}[b]{\linewidth}
        \centering
        \includegraphics[width=0.92\textwidth, height=1.885cm]{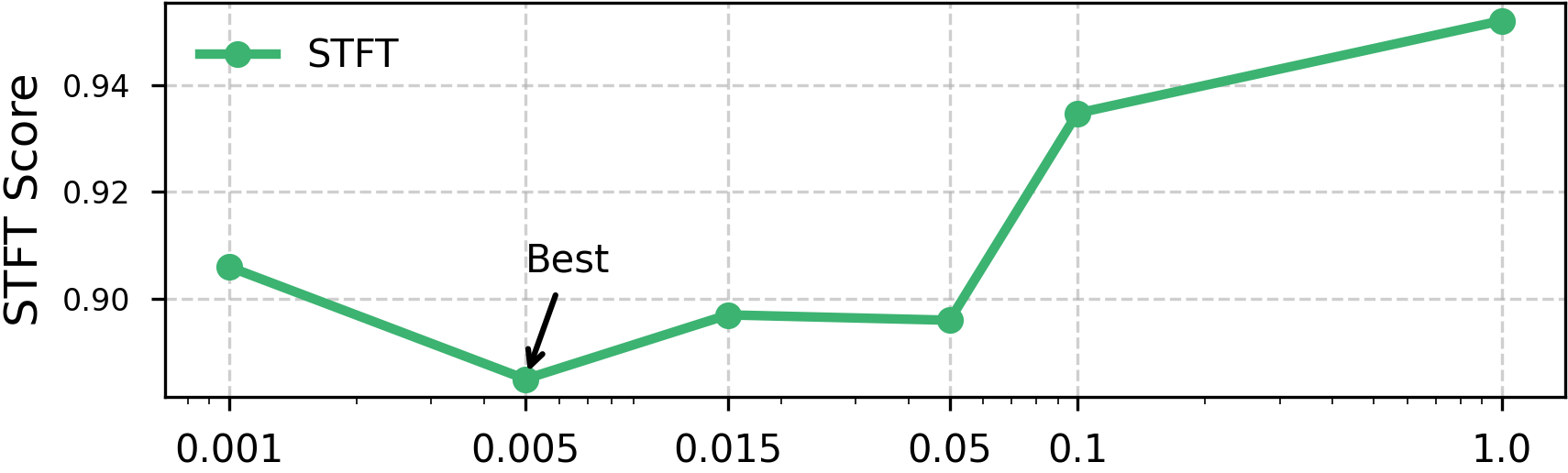}
        \vspace{-6pt}
        \caption{Hyper-parameter $\zeta$}
        \label{fig:ablation_zeta}
    \end{subfigure}
    \begin{subfigure}[t]{\linewidth}
        \centering
        \includegraphics[width=0.92\textwidth, height=1.885cm]{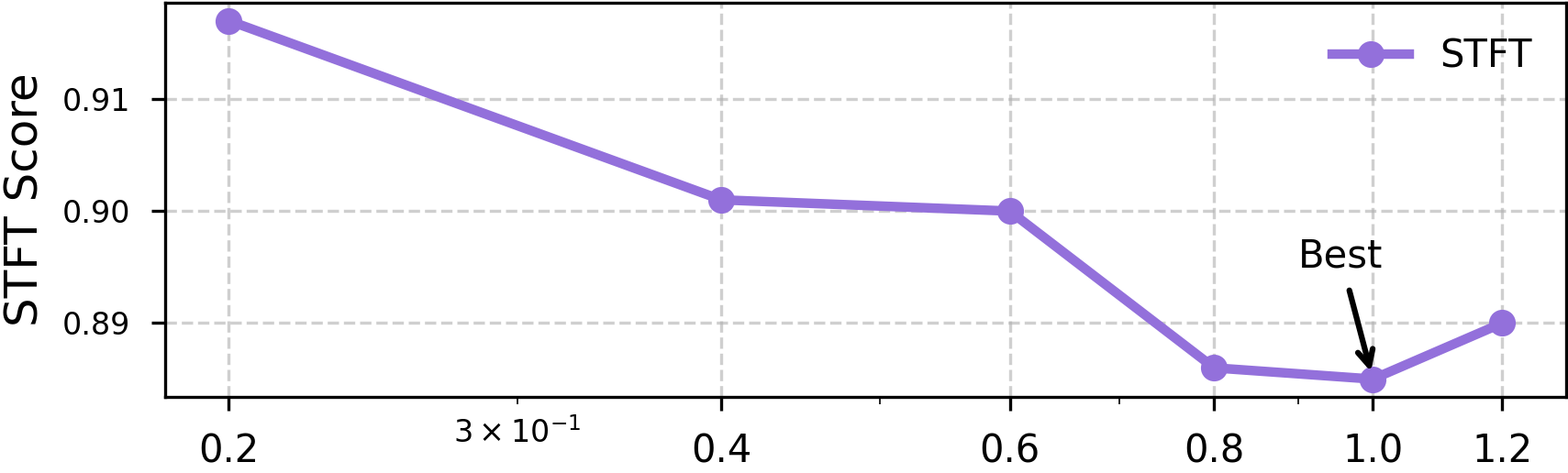}
        \vspace{-6pt}
        \caption{Hyper-parameter $\eta$}
        \vspace{-10pt}
        \label{fig:ablation_eta}
    \end{subfigure}
    \caption{
    Ablation study of the model hyper-parameters $\lambda$, $\zeta$, and $\eta$ on the FAIR-Play dataset (5 splits)~\cite{gao20192,xu2021visually}, evaluated using the $\text{STFT} \downarrow$ metric.
    }
    \vspace{-15pt}
    \label{fig:ablation_hyper}
    \Description[Ablation study the model hyper-parameters]{Ablation study the model hyper-parameters}
\end{figure}

\subsection{Results}
\noindent \textbf{Results on FAIR-Play Dataset.}
We adopt established benchmarks for conducting model evaluations, such as FAIR-Play 10-splits~\cite{gao20192} and 5-splits~\cite{xu2021visually}. We first show the comparison on FAIR-Play 10-split~\cite{gao20192} benchmark in Tab.~\ref{tab:fairplay_10s}. 
The results demonstrate that our model surpasses the second-best models with a \textit{relative improvement} of $+3.01\%$ in $\text{STFT}$, $+0.89\%$ in $\text{ENV}$ and $3.20\%$ in $\text{WAV}$, respectively.
Please note that we exclude CLUP~\cite{li2024cyclic} from this table, as it introduces additional computational complexity (e.g., diffusion~\cite{ho2020denoising} and VGGish~\cite{hershey2017cnn}), and their method is not publicly available for inference comparison.
To evaluate the true generalisation ability as suggested by PseudoBinaural~\cite{xu2021visually}, we also utilise the newly proposed FAIR-Play (5-split)~\cite{xu2021visually} for the evaluation, as shown in Tab.~\ref{tab:fairplay_5s}. 
We re-implemented CMC~\cite{liu2024visually} for the FAIR-Play (5-split) benchmark, as the original paper did not report results under this setting. Since the official implementation of CMC was not publicly available at the time, we re-implemented the model based on the details provided in the paper.
Our method outperforms the second-best model with a \textit{relative improvement} of
$+2.54\%$ in $\text{STFT}$, $+0.73\%$ in $\text{ENV}$, $+2.48\%$ in $\text{Mag}$, $1.69\%$ in $\text{Phs}$, and $+1.22\%$ in $\text{SNR}$.
Please note that all reported metrics (e.g., STFT) are challenging to improve, as the STFT provides a high time-frequency resolution, making differences less significant than other metrics like ROC-AUC score.

\noindent \textbf{Results on Real-world YouTube-based datasets.}
To further evaluate model scalability and generalisability on larger-scale real-world datasets, we follow~\cite{xu2021visually,parida2022beyond} to assess performance on the MUSIC-Stereo dataset~\cite{xu2021visually} and YT-MUSIC~\cite{morgado2018self}, as shown in Tab.~\ref{tab:music_stereo} and Tab.~\ref{tab:yt_music}. 
Our method outperforms the second-best model with a \textit{relative improvement} of
$+6.87\%$ in $\text{STFT}$, $+10.19\%$ in $\text{ENV}$, $+6.87\%$ in $\text{Mag}$, $+2.34\%$ in $\text{Phs}$, and $+20.75\%$ in $\text{SNR}$ on the MUSIC-Stereo dataset~\cite{zhao2018sound,xu2021visually}, and $+5.70\%$ in $\text{STFT}$, $+0.70\%$ in $\text{ENV}$, $+11.40\%$ in $\text{Mag}$, $+6.80\%$ in $\text{Phs}$, and $+63.50\%$ in $\text{SNR}$ on the YT-MUSIC dataset~\cite{morgado2018self,xu2021visually}.

\subsection{Ablation Study}
\noindent \textbf{Ablation of Key Components.}
We perform an analysis of CCStereo components on the second split of FAIR-Play (5-split)~\cite{xu2021visually}, as shown in Tab.~\ref{tab:ablation_components}. Starting from a baseline (1st row) consisting of a simple U-Net model similar to Mono2Binaural~\cite{gao20192} that resizes the input frame directly to 448$\times$224 during the inference. We utilise the TDSS method to enhance the inference process, resulting in an $\text{STFT}$ improvement of $+2.49\%$. Integrating AVAD into the system (3rd row) provides an additional improvement of $+1.03\%$. Subsequently, adding $\ell_{\text{AMP}}$ and $\ell_{\text{PHS}}$ (i.e., $\ell_{\text{REC}}$) (4th row) and incorporating the $\ell_{\text{SCL}}$ contrastive learning method (5th row) yield further improvements of $+1.80\%$ and $+0.65\%$, respectively.

\noindent \textbf{Ablation of the Reconstruction Loss.}
To separately analyse each loss term in $\ell_{\text{REC}}$, we conducted an ablation study, as shown in Fig.~\ref{fig:ablation_rec_loss}, to evaluate the individual contributions of $\ell_{\text{AMP}}$ and $\ell_{\text{PHS}}$. Starting with the third row in Tab.~\ref{tab:ablation_components}, we progressively added $\ell_{\text{AMP}}$ and $\ell_{\text{PHS}}$ during model training. 
We observed improvements of $+0.41\%$ and $+1.40\%$ on STFT, respectively, highlighting the importance of aligning phase information for accurate binaural prediction. 
To better demonstrate the importance of the $\ell_{\text{AMP}}$ and $\ell_{\text{PHS}}$ losses for the binaural audio generation task, we provide a qualitative visualisation of the predicted real spectrograms. Fig.~\ref{fig:ablation_without_rec} shows the results using only $\ell_{\text{MSE}}$, while Fig.~\ref{fig:ablation_with_rec} uses $\ell_{\text{REC}}$, which is a combination of $\ell_{\text{MSE}}$, $\ell_{\text{AMP}}$, and $\ell_{\text{PHS}}$. The top row displays the ground truth (GT) real spectrogram, the middle row shows the predicted spectrogram, and the bottom row illustrates the point-wise MSE. 
Compared to Fig.~\ref{fig:ablation_without_rec}, the model trained with the combined loss in Fig.~\ref{fig:ablation_with_rec} produces a prediction that is visually more aligned with the ground truth, with lower reconstruction error, particularly in fine-grained high-frequency details. This highlights the complementary role of amplitude and phase-aware losses in improving perceptual quality.

\noindent \textbf{Ablation of the Spatial-aware Contrastive Learning}
We conducted an ablation study on the contrastive loss weight $\lambda$, as illustrated in Fig.~\ref{fig:ablation_lambda}. The results suggest that assigning a large value to $\lambda$ causes the contrastive loss to dominate the primary MSE loss, potentially hindering the model's ability to optimise for the core BAG objective. In contrast, using a smaller $\lambda$ helps maintain a balance between representation learning and the main training objective, enabling effective structuring of the pixel embedding space without compromising BAG performance. A similar trade-off was also discussed in~\cite{zhou2022rethinking}.

\noindent \textbf{Hyper-Parameters Analysis}
We conduct an ablation study to investigate the sensitivity of the hyper-parameters on the FAIR-Play (5-split)~\cite{gao20192,xu2021visually} dataset, as shown in Fig.~\ref{fig:ablation_zeta} and Fig.~\ref{fig:ablation_eta}. The results indicate that it is necessary to weight $\ell_{\text{AMP}}$ and $\ell_{\text{PHS}}$ differently: the best performance is achieved with a small value for $\zeta$, while $\eta$ stabilises around 1.0. We attribute this to two factors: (1) the amplitude loss typically has a much larger magnitude than phase-related losses, and (2) perceptual quality in binaural audio strongly depends on both amplitude and phase. If the model is already struggling to learn accurate phase information, increasing the emphasis on amplitude (via a larger $\zeta$) may suppress phase learning and lead to "fuzzy" or "muffled" outputs. These findings highlight the importance of carefully balancing the two objectives during training.

\subsection{Qualitative Results}
We present qualitative results of visual activation estimated by our method in Fig.~\ref{fig:attn_visual}. Specifically, we extract the output from the convolution layer for Mono2Binaural~\cite{gao20192}, Sep-Stereo~\cite{zhou2020sep} and AVAD, average the activation map across channels, and normalise it using min-max normalisation. The results in Fig.~\ref{fig:attn_visual}\textcolor{custompurple}{a}, ~\ref{fig:attn_visual}\textcolor{custompurple}{b} indicate that our method 
can better focus on the sounding object and its position. 
\revise{
In some cases, when the instrument is not clearly detected, as shown in Fig.\ref{fig:attn_visual}\textcolor{custompurple}{c}, the model instead shifts its attention to the performer’s motion. 
}
However, Tab.~\ref{fig:attn_visual}\textcolor{custompurple}{d} illustrates a failure case, where the model is unable to localise the occluded object ``piano'' and instead shows a tendency to focus on the room environment (Room Env).
We hypothesise that when the model fails to identify the sounding object, it associates the audio with the room environment \revise{as these environmental cues provide a more consistent and easily exploitable shortcut signal.} These observations highlight the limitations of the current method.
For further results on videos, refer to the \textit{Supplementary Material}.

\section{Discussion and Conclusion}
We introduced CCStereo, a new audio-visual training method designed for the U-Net-based framework to enhance spatial awareness and reduce overfitting to room environments. 
We proposed a visually conditioned adaptive de-normalisation method that utilises the object's spatial information to guide the decoding of the difference audio. To enhance the representation learning of spatial awareness, we design a new audio-visual contrastive learning based on mining
negative samples from randomly shuffled visual feature
representation. Furthermore, our cost-efficient 
test-time dynamic scene simulation 
strategy enhanced robustness without adding computational overhead.
Our approach consistently outperformed existing methods on the FAIR-Play, MUSIC-Stereo and YT-MUSIC datasets, achieving state-of-the-art results across various metrics.

\section*{Acknowledgements}
We sincerely thank Toshimitsu Uesaka for their valuable feedback and insightful suggestions. 
Yuanhong Chen acknowledge financial support from Commonwealth Bank of Australia under the CommBank Centre for Foundational AI Research. The collaboration facilitated by this funding has significantly contributed to the progress of this research.

\appendix
\section{Sliding Window Integration}
\begin{figure}[tp]
    \centering
    \includegraphics[width=.9\linewidth]{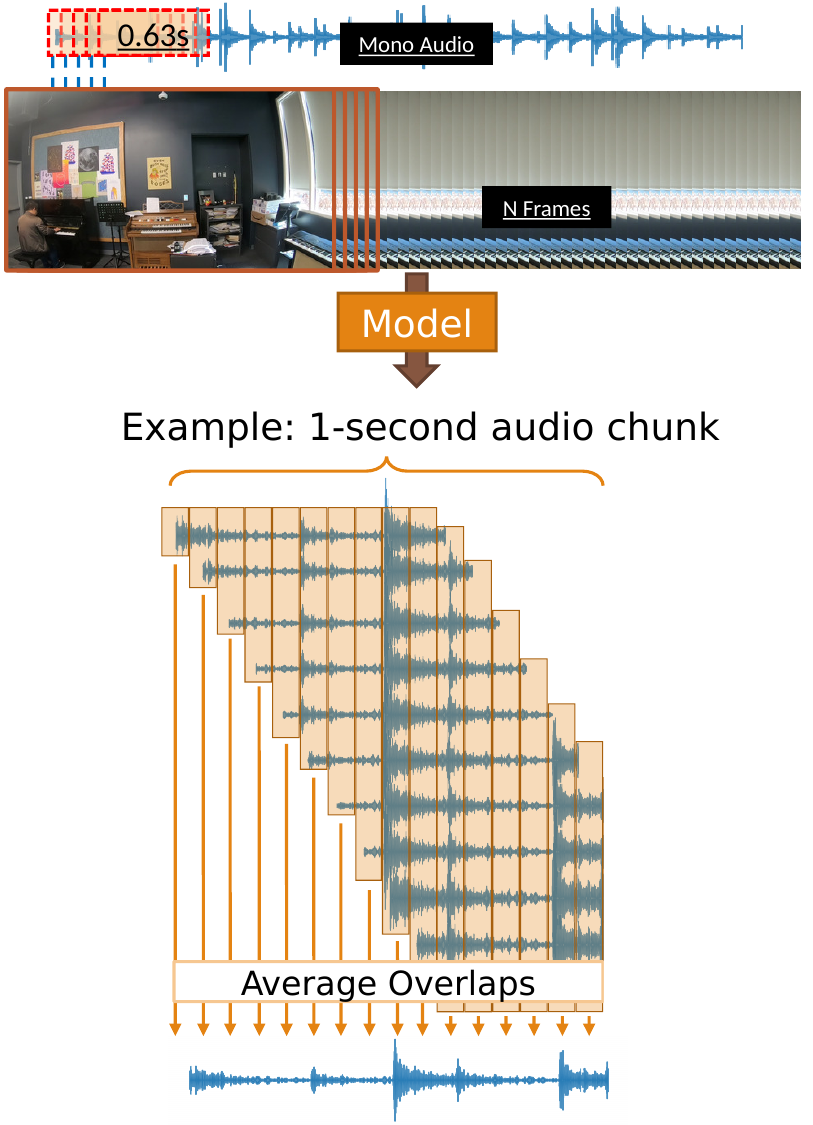}
    \vspace{-10pt}
    \caption{Illustration of sliding window integration in Mono2Binaural~\cite{gao20192}.}
    \label{fig:sliding_windows_integ}
\end{figure}

We adopt the sliding window integration method from Mono 2Binaural~\cite{gao20192} during model inference, as shown in Fig.~\ref{fig:sliding_windows_integ}, to enable the model to handle moving sound sources and camera motion~\cite{gao20192}. The input monaural audio is divided into $N$ audio segments with a hop size of 0.1, where each segment corresponds to a video frame. After predicting each audio segment, the predicted audio chunks are integrated by averaging their overlapping predictions to form the final difference audio prediction.

\begin{table}[tp]
    \centering
    \def\arraystretch{1.1}
    \caption{Comparison with existing approaches on \textbf{YT-CLEAN}~\cite{morgado2018self}. Best results are shown in \textbf{bold}, and the $2^{nd}$ best are \underline{underlined}.}
    \vspace{-10pt}
    \resizebox{0.7\linewidth}{!}{
    \begin{tabular}{!{\vrule width 1.1pt}l!{\vrule width 1.1pt}cc!{\vrule width 1.1pt}}
    \specialrule{1.1pt}{0pt}{0pt}
    \multirow{2}{*}{Methods} & \multicolumn{2}{c!{\vrule width 1.1pt}}{YT-CLEAN~\cite{morgado2018self}} \\
    \cline{2-3}
    ~ & $\text{STFT} \downarrow$ & $\text{ENV} \downarrow$ \\
    \specialrule{1.1pt}{0pt}{0pt}
    Ambisonics~\cite{morgado2018self} & 1.435 & 0.155 \\
    Mono-Mono & 1.407 & 0.141 \\
    Mono2Binaural~\cite{gao20192, xu2021visually} & \underline{1.073} & \underline{0.133} \\
    \hline
    \rowcolor{LightCyan}\textbf{CCStereo} & \textbf{0.944} & \textbf{0.125} \\
    \specialrule{1.1pt}{0pt}{0pt}
    \end{tabular}
    }
    \vspace{-8pt}
    \label{tab:yt_clean}
\end{table}

\section{Additional Results}
We acknowledge the importance of evaluating model performance on audio-visual content captured in natural, unconstrained environments. To this end, we conducted supplementary evaluations on the YT-CLEAN dataset~\cite{morgado2018self}, which comprises in-the-wild audio-visual recordings. Compared to curated musical content, this dataset presents a more diverse and challenging setting, providing valuable insights into a model’s ability to generalise.
As shown in Table~\ref{tab:yt_clean}, our method \textbf{CCStereo} achieves the best performance on both the STFT and ENV metrics, outperforming the Mono2Binaural baseline by 12.02\% and 6.02\%, respectively. These results underscore the limitations of existing approaches when applied to complex, less structured real-world scenes and demonstrate the robustness of our method in such conditions.

\section{Qualitative Results}
\begin{figure*}[ht]
    \centering
    \begin{subfigure}[b]{.49\linewidth}
         \centering    
            \includegraphics[width=1.\linewidth]{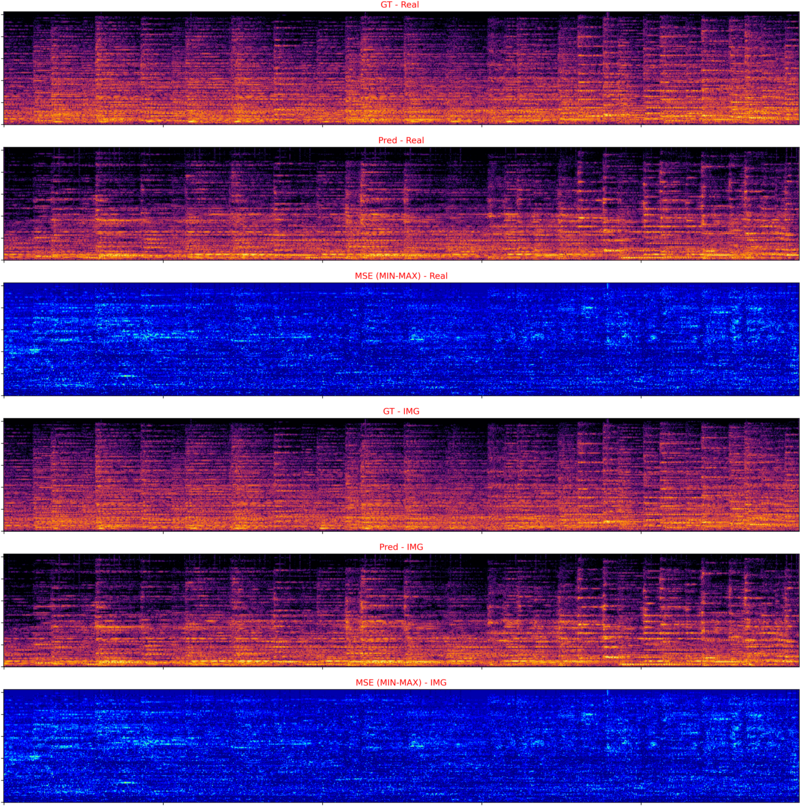}
        \caption{Mono2Binaural~\cite{gao20192}}
    \end{subfigure}
    \begin{subfigure}[b]{.49\linewidth}
        \centering    
            \includegraphics[width=1.\linewidth]{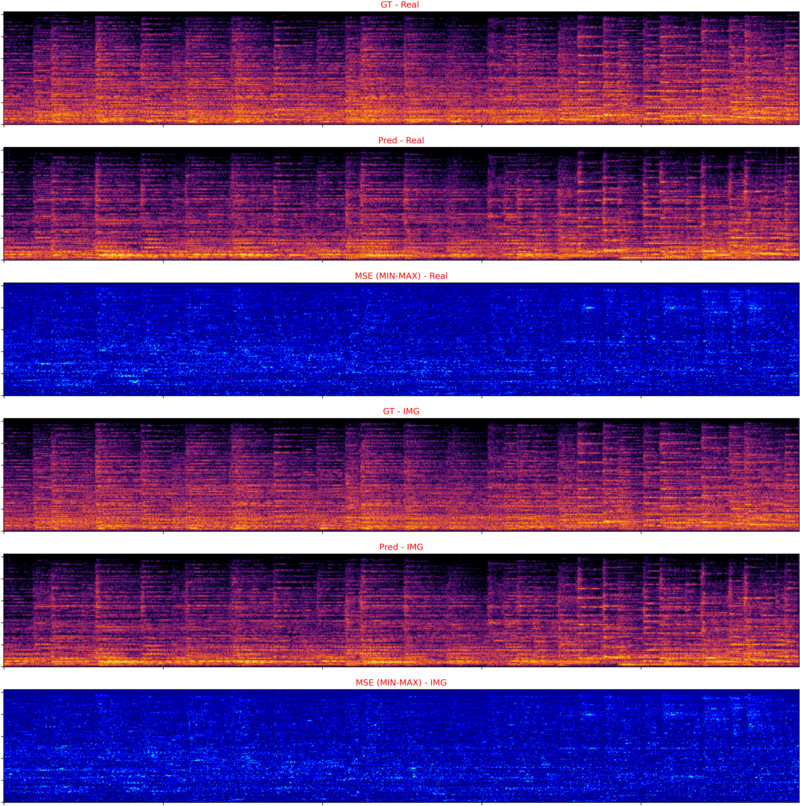}
        \caption{CCStereo}
    \end{subfigure}
    \\
    \begin{subfigure}[b]{.49\linewidth}
         \centering    
            \includegraphics[width=1.\linewidth]{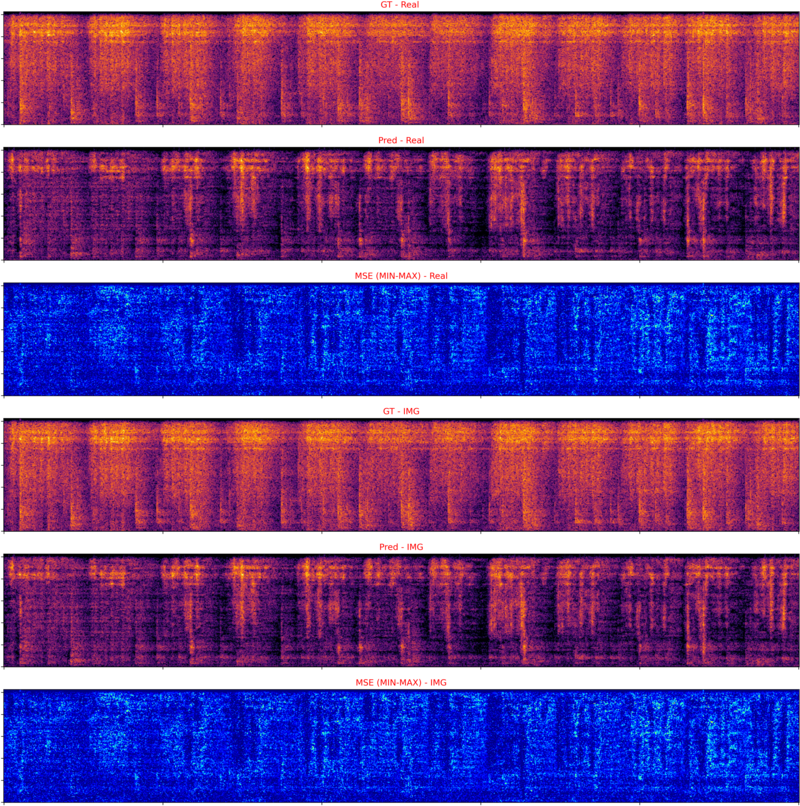}
        \caption{Mono2Binaural~\cite{gao20192}}
    \end{subfigure}
    \begin{subfigure}[b]{.49\linewidth}
        \centering    
            \includegraphics[width=1.\linewidth]{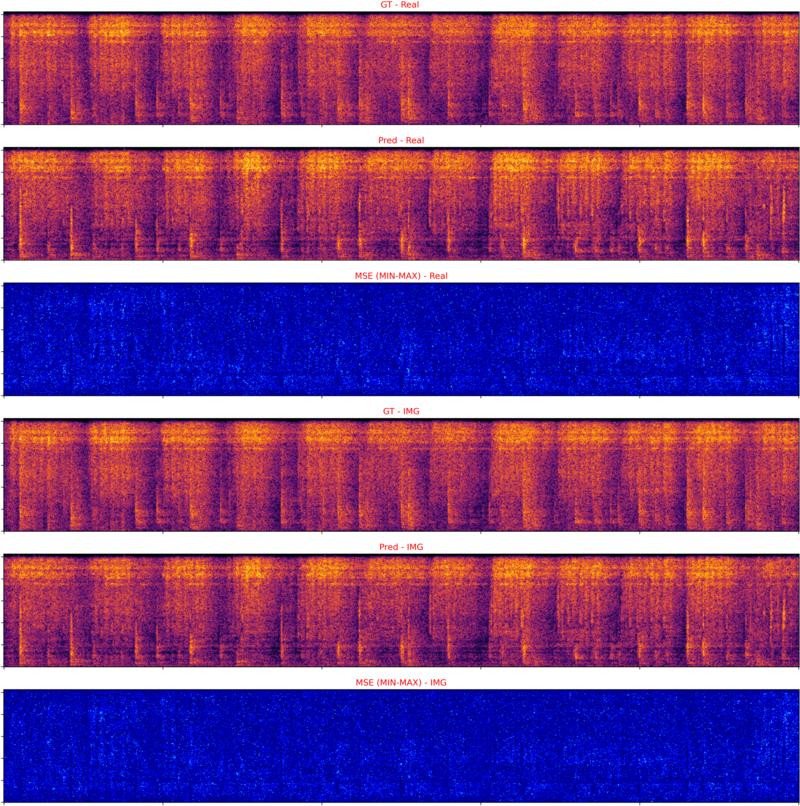}
        \caption{CCStereo}
    \end{subfigure}
    \caption{Qualitative results on FAIR-Play dataset~\cite{gao20192}.}
    \label{fig:visual_3}
\end{figure*}

We present a qualitative comparison visualisation between Mono 2Binaural~\cite{gao20192} and our proposed CCStereo in Fig.~\ref{fig:visual_3}. The spectrogram of the ground-truth difference audio is shown in the \nth{1} (real) and \nth{4} (imaginary) rows, while the predictions of each method are displayed in the \nth{2} and \nth{5} rows. Additionally, we provide the mean square error (MSE) results in the \nth{3} and \nth{6} rows to highlight prediction accuracy. These findings demonstrate that our method approximates the true difference in audio more accurately, showcasing the effectiveness of our approach.

\bibliographystyle{ACM-Reference-Format}
\balance
\bibliography{egbib}
\end{document}